\documentclass{aastex}
\usepackage{emulateapj5}
\usepackage{apjfonts}
\usepackage{epsf}

\slugcomment{UTAP-448}
\shorttitle{DECAYING COLD DARK MATTER}
\shortauthors{OGURI ET AL.}
\begin{document}
\title{Decaying Cold Dark Matter and the Evolution of the Cluster Abundance}
%
\author{Masamune Oguri\altaffilmark{1}, Keitaro Takahashi\altaffilmark{1}, Hiroshi Ohno\altaffilmark{2,3} and Kei Kotake\altaffilmark{1}}
\altaffiltext{1}{Department of Physics, School of Science, University of Tokyo, Hongo 7-3-1, Bunkyo-ku, Tokyo 113-0033, Japan}
\altaffiltext{2}{Division of Theoretical Astrophysics, National
Astronomical Observatory, 2-21-1 Osawa, Mitaka, Tokyo 181-8588, Japan}
\altaffiltext{3}{Research Center for the Early Universe (RESCEU), School
of Science, University of Tokyo, Hongo 7-3-1, Bunkyo-ku, Tokyo 113-0033, Japan}
\email{oguri@utap.phys.s.u-tokyo.ac.jp, ktaro@utap.phys.s.u-tokyo.ac.jp,
ohno@resceu.s.u-tokyo.ac.jp, kkotake@utap.phys.s.u-tokyo.ac.jp}
%
%
\begin{abstract}
 The cluster abundance and its redshift evolution are known to be a
 powerful tool to constrain the amplitude of mass fluctuations $\sigma_8$
 and the mass density parameter $\Omega_{m0}$. We study the impact of
 the finite decay rate of cold dark matter particles on the cluster
 abundances. On the basis of spherical model in decaying cold dark
 matter universe, we calculate the mass function of clusters and compare
 it with observed cluster abundance. We find the decay of cold dark
 matter particles significantly changes the evolution of the cluster
 abundance. In particular, we point out that the lifetime of dark matter
 particles comparable to the age of the universe lowers the ratio of the
 local cluster abundance to the high-redshift cluster abundance and can
 account for the observed evolution of the cluster abundance quite well.
 The strong dependence of the cluster abundance on the decay rate of
 dark matter suggests that distant cluster surveys may offer clues to
 the nature of dark matter. 
\end{abstract}  
%
\section{Introduction}
The abundance of cluster of galaxies places a strong constraint on
fundamental cosmological parameters such as the amplitude of 
mass fluctuations on a scale of $8h^{-1}{\rm Mpc}$, $\sigma_8$, and the
present mass density parameter, $\Omega_{m0}$
\citep*{henry91,bahcall92,white93,eke96,kitayama97}. Although the
abundance of clusters is strong function of both $\Omega_{m0}$ and
$\sigma_8$, the evolution of the cluster abundance with redshift breaks
the degeneracy between $\Omega_{m0}$ and $\sigma_8$ 
\citep{eke96,henry97,henry00,bahcall98,bahcall03b}.
For instance, \citet{bahcall03b} pointed out that the relatively high
abundance of massive cluster at $z>0.5$ prefers large mass fluctuations 
$\sigma_8=0.9-1$. On the other hand, the local cluster abundance in both
optical \citep{bahcall03a} and X-ray \citep{ikebe02,seljak02} wavebands
suggests low mass fluctuations $\sigma_8=0.7-0.8$ for
$\Omega_{m0}\sim 0.3$. The combination of local and high-$z$
cluster abundances yields relatively low matter density in the universe,
e.g., $\Omega_{m0}=0.17\pm0.05$ \citep{bahcall03b}. However, other
observations, such as the anisotropy of cosmic microwave background
\citep{spergel03} and Type Ia supernovae \citep{tonry03}, prefer
somewhat larger matter density; $\Omega_{m0}=0.27\pm 0.04$ from WMAP plus
large scale structure \citep{spergel03} or $\Omega_{m0}=0.28\pm0.05$
from Type Ia supernovae \citep{tonry03}. Although all
these results are consistent with each other within $2\sigma$ level,
the slight difference of $\sigma_8$ or $\Omega_{m0}$ might be
interpreted that underlying models are wrong \citep{bridle03}. 

Although the above constraints are derived assuming the usual stable cold
dark matter model, the evolution may be considerably different
if cold dark matter particles are not perfectly stable and decay into
relativistic particles with a decay rate $\Gamma$ \citep{cen01}. 
Therefore, the evolution of the cluster abundance may become a useful
test to explore the nature of dark matter. 
The decaying cold dark matter model has been studied so far in several
contexts, e.g., to reconcile Einstein-de Sitter universe with current
low matter universe \citep*{turner84}, to explain unexpected X-ray
observations \citep{dilella03}, or to reduce possible over-concentration
of dark matter and over-abundance of substructures in cold dark matter
model \citep{cen01,sanchez03}. Moreover, cold dark matter with similar
property, disappearing dark matter, has also attracted much attention in
the context of the brane world scenario 
\citep{randall99a,randall99b}. In the brane world scenario a bulk scalar
field can be trapped on the brane, which is identified with our universe,
and become cold dark matter. But since the scalar field is expected to be
metastable, it decays into continuum states in the higher dimension with
some decay width $\Gamma$, which is determined by the mass of the scalar 
field and the energy scale of the extradimension \citep*{dubovsky00}. 
An observer on the brane sees this as if the scalar field disappears into 
the extradimension leaving some energy on the brane. This energy behaves 
in the same way as energy of relativistic particles but does not
correspond to real particles. Thus this is called ``dark radiation''. 
Although the disappearing dark matter and the decaying dark matter are
not the same, they contribute to the expansion of the universe in the same
way. Hereafter we call both of them decaying dark matter.
\citet{ichiki03} studied the universe with the decaying dark matter as a
test of brane world scenario. They found that decaying cold dark matter
model with lifetime about the age of the universe improve the fit of
Type Ia supernovae observations and the evolution of mass-to-light
ratios of clusters.   

In this paper, we study structure formation in the decaying cold 
dark matter universe; we present the spherical model, make theoretical
predictions for the cluster abundance, and show that the evolution of
the cluster abundance is quite sensitive to the decay rate of dark
matter particles.  We also compare our predictions with observed low-$z$
and high-$z$ cluster abundances, and see if the decaying cold dark
matter model explains the relatively large cluster abundance at high-$z$
universe. 

\section{Theoretical Model}

To study the cluster abundance, we need the mass function of dark halos
in decaying cold dark matter model. In this section, we construct the
mass function using the spherical model as in \citet{press74}. We also
present the nonlinear overdensity $\Delta_{\rm c}$ which is important in
studying the cluster abundance. Throughout this paper we assume a flat
universe.  

\subsection{Cosmology and Linear Growth Rate}

We consider the case that dark matter particles decay into
relativistic particles, which are assumed to contribute to the structure
formation only through the change of the expansion law of the universe
by their energy behavior. Rate equations of matter and radiation
components in the decaying dark matter model are
$\dot{\rho}_m+3H\rho_m=-\Gamma\rho_m$ and
$\dot{\rho}_r+4H\rho_r=\Gamma\rho_m$. This results in the following
evolution of matters and radiations: $\rho_m=\rho_{m0}a^{-3}e^{-\Gamma
t}$ and $\rho_r=\Gamma\rho_{m0}a^{-4}\int_0^tae^{-\Gamma t}dt$, respectively.
Here we assume no radiation component except for decay products. The
Friedmann equation then becomes
\begin{equation}
\frac{H^2(a)}{H_0^2}=\Omega_{m0}a^{-3}e^{-\Gamma(t-t_0)}+\Gamma\Omega_{m
0}a^{-4}\int_0^tae^{-\Gamma(t-t_0)}dt+\lambda_0.
\label{ea}
\end{equation}
This equation should be solved with the equation of the cosmological
time $t=\int_0^a da/(aH(a))$ simultaneously. By combining these two
equations, we obtain a following differential equation
\begin{equation}
 t''=-\frac{H_0^2}{2}\left(\frac{\Omega_{m0}e^{-\Gamma(t-t_0)}}{a^2}+4a\lambda_0\right)t'{}^3+\frac{t'}{a},
\end{equation}
where $'\equiv d/da$. Note that the current density parameter
$\Omega_{m0}$ should satisfy the Freidmann equation at $z=0$:
\begin{equation}
 1=\Omega_{m0}+\Gamma\Omega_{m0}\int_0^{t_0}ae^{-\Gamma(t-t_0)}dt+\lambda_0.
\end{equation} 
Therefore the current density parameter $\Omega_{m0}$ is uniquely
determined for given $\Gamma$ and $\lambda_0$, i.e.,
$\Omega_{m0}=\Omega_{m0}(\Gamma,\lambda_0)$. 
Next consider the linear perturbation. The density perturbation is
denoted by $\delta_m\equiv (\rho_m-\bar{\rho}_m)/\bar{\rho}_m$, where
$\bar{\rho}_m$ is background matter density. Then perturbation equation becomes
\begin{equation}
 \dot{a}^2\delta_m''+\left(\ddot{a}+2\frac{\dot{a}^2}{a}\right)\delta_m'-4\pi G\bar{\rho}_m\delta_m=0.
\label{linear}
\end{equation}
We solve this differential equation with the boundary condition
$\delta_m\propto a$ at $a\ll 1$.

\subsection{Spherical Model}

In this subsection, we describe the spherical collapse model in decaying
dark matter model and give the methodology to calculate nonlinear
overdensity $\Delta_{\rm c}$. First we consider a spherical overdensity
with initial mass $M$ and radius $R$. The equation of motion of such a
spherical shell is given by
\begin{equation}
 \frac{\ddot{R}}{R}=-\frac{GM}{R^3}e^{-\Gamma t}-H_0^2\Gamma\Omega_{m0}a^{-4}\int_0^tae^{-\Gamma(t-t_0)}dt+H_0^2\lambda_0.
\end{equation}
We define the following quantities: $x\equiv a/a_{\rm ta}$, $y\equiv
R/R_{\rm ta}$, $\tau\equiv H(x=1)\sqrt{\Omega_m(x=1)}t$,
$\gamma\equiv\Gamma/(H(x=1)\sqrt{\Omega_m(x=1)})$, and
$\eta\equiv\lambda(x=1)/\Omega(x=1)$. Here $a_{\rm ta}$ represents the scale
factor at turnaround. Then equations we should solve are
the two dimensionless equations:
\begin{equation}
 \frac{d^2\tau}{dx^2}=-\frac{1}{2}\left(\frac{e^{-\gamma(\tau-\tau_{\rm ta})}}{x^2}+4\eta x\right)\left(\frac{d\tau}{dx}\right)^3+\frac{1}{x}\frac{d\tau}{dx},
  \label{time_sc}
\end{equation}
\begin{eqnarray}
\frac{d^2y}{dx^2}=\frac{d^2\tau}{dx^2}\left(\frac{d\tau}{dx}\right)^{-1}\frac{dy}{dx}+\left(\frac{d\tau}{dx}\right)^2&&\nonumber\\
&&\hspace*{-50mm}\times\left[-\frac{1}{2}y^{-2}\zeta e^{-\gamma(\tau-\tau_{\rm ta})}-\gamma yx^{-4}\int_0^\tau xe^{-\gamma(\tau-\tau_{\rm ta})}d\tau +\eta y\right],
\label{radius_sc}
\end{eqnarray}
where $\zeta$ is the density contrast of the spherical overdensity
region at turnaround, $\zeta\equiv \rho_{\rm
spherical}/\bar{\rho}_m|_{x=1}$. We solve these equation from $\tau=0$
to the virialization of the overdensity region with boundary conditions
$\tau(x\ll 1)\simeq (2/3)e^{-\gamma\tau_{\rm ta}/2}x^{3/2}$,
$\tau(x=1)=\tau_{\rm ta}$, $y(x=0)=0$, $y(x=1)=1$, and $dy/dx(x=1)=0$ in
order to derive $\tau_{\rm ta}$ and $\zeta$. Although the
conventional choice of the virialization epoch is the collapse time
($y=0$), in this paper we define the virialization epoch when the virial
theorem in the following holds: 
\begin{equation}
 \left(\frac{dy}{d\tau}\right)^2=\frac{1}{2}y^{-1}\zeta e^{-\gamma(\tau-\tau_{\rm ta})}-\gamma y^2 x^{-4}\int_0^\tau xe^{-\gamma(\tau-\tau_{\rm ta})}d\tau +\eta y^2.
\end{equation}
Given the virialization epoch and virial radius, the nonlinear
overdensity can be calculated from $\Delta_{\rm c}(x_{\rm vir})=(x_{\rm
vir}^3/y_{\rm vir}^3)\zeta-1$. To calculate the cluster abundance, we
also need the extrapolation of the linear fluctuation $\delta_{\rm c}$
at virialization.  We use $\delta_{\rm c}=1.58$, which is the
value in Einstein-de Sitter universe for all cosmological models, mainly
because of the  computational cost. This simplification does not change
our results because we confirmed that $\delta_{\rm c}$ only weakly
depends on the cosmological parameters. 

\subsection{Mass Function}

We derive the mass function of clusters using the theory of
\citet{press74}. First start from an initial density field
$\delta(\vec{x}, M_i, z_i)$ smoothed over the region containing mass
$M_i$. If the initial density field is random Gaussian, a probability
distribution function of $\delta$ at any point is given by   
\begin{equation}
P\left[\delta(M_i, z_i)\right]=\frac{1}{\left(2\pi\right)^{1/2}\sigma_{M_i}(z_i)} \exp\left[-\frac{\delta^2(M_i, z_i)}{2\sigma^2_{M_i}(z_i)}\right],
\end{equation}
where $\sigma_{M_i}(z_i)=\sigma(R_{M_i}, z_i)$ is the mass variance.
 From the spherical model, it can be interpreted that the region is
already virialized at $z$ if the linearly extrapolated density contrast
$\delta_{\rm linear}(M_i, z)$ exceeds the critical value $\delta_{\rm
c}$. Therefore the probability that the region with mass $M$ is already
virialized is given by
\begin{equation}
 f(M, t)=\frac{1}{2}{\rm erfc}\left(\frac{\delta_{\rm c}(z)}{\sqrt{2}\sigma_{M_i}}\right),
\label{fractionfunction}
\end{equation}
where ${\rm erfc}(x)$ is the complementary error function, $\delta_{\rm
c}(z)\equiv\delta_{\rm c}D(z=0)/D(z)$, and
$\sigma_{M_i}\equiv\sigma_{M_i}(z=0)$, where $D(z)$ is linear growth
rate calculated from equation (\ref{linear}). We assume the mass
variance for the cold dark matter fluctuation spectrum with the
primordial spectral index $n=1$, and adopt a fitting formula presented
by \citet{kitayama96}. From equation (\ref{fractionfunction}), we
finally obtain the comoving  number density of halos of mass $M$ at time
$z$: 
\begin{equation}
 \frac{dn_{\rm PS}}{dM}(M, z)=\left.e^{2\Gamma t}\sqrt{\frac{2}{\pi}}\frac{\rho_0}{M_i}\frac{\delta_{\rm c}(z)}{\sigma_{M_i}^2}\left|\frac{d\sigma_{M_i}}{dM_i}\right|\exp\left[-\frac{\delta_{\rm c}^2(z)}{2\sigma_{M_i}^2}\right]\right|_{M_i=Me^{\Gamma t}},
\label{pressschechter}
\end{equation}
with $\rho_0=\rho_{\rm crit}(z=0)\Omega_{m0}e^{-\Gamma(t-t_0)}$.

\section{Evolution of the Cluster Abundance}
In the model described in the previous section, the abundance and its
redshift evolution of clusters are fully determined by three independent
parameters; $\lambda_0$, $\sigma_8$, and $\Gamma$. In this section, we
see the dependence of the cluster abundances on these parameters.

Figure \ref{fig:cluabu} plots the number density of clusters as a
function of redshift. We consider the cases with the lifetime of dark matter
$\Gamma^{-1}=10h^{-1}{\rm Gyr}$, and also with the usual infinite
lifetime of dark matter for reference. Following \citet{bahcall03b}, we
adopt the mass within a comoving radius of $1.5h^{-1}{\rm Mpc}$,
$M_{1.5h^{-1}{\rm Mpc}}$, to compare our theoretical predictions with
observed cluster abundance. For the extrapolation to  $1.5h^{-1}{\rm
Mpc}$, we use the observed cluster profile, $M(<R)\sim R^{0.6}$
\citep*{carlberg97,fischer97}. For the observed cluster abundance, we
plot the data compiled by \citet{bahcall03b}; abundance at $z\sim 0.05$
from \citet{ikebe02}, at $z\sim 0.38$ from \citet{henry00}, at $z\sim
0.5-0.65$ and $z\sim 0.65-0.9$ from \citet{bahcall98}. Obviously, inclusion of
finite lifetime greatly improves the situation: for
$\Gamma^{-1}=10h^{-1}{\rm Gyr}$ both local and high-$z$ cluster
abundance can be reproduced with the same $\sigma_8$. For
$\Gamma^{-1}=\infty$ it is difficult to explain local ($z\sim 0.05$) and
high-$z$ ($z\sim 0.3-0.9$) cluster abundances simultaneously, although
both can be reconciled if we adopt larger $\lambda_0$,
$\lambda_0\sim0.8$. This drastic change of the evolution of the cluster
abundance implies that it can place a strong constraint on the decay
rate of cold dark matter. 

\vspace{0.5cm}
\centerline{{\vbox{\epsfxsize=7.5cm\epsfbox{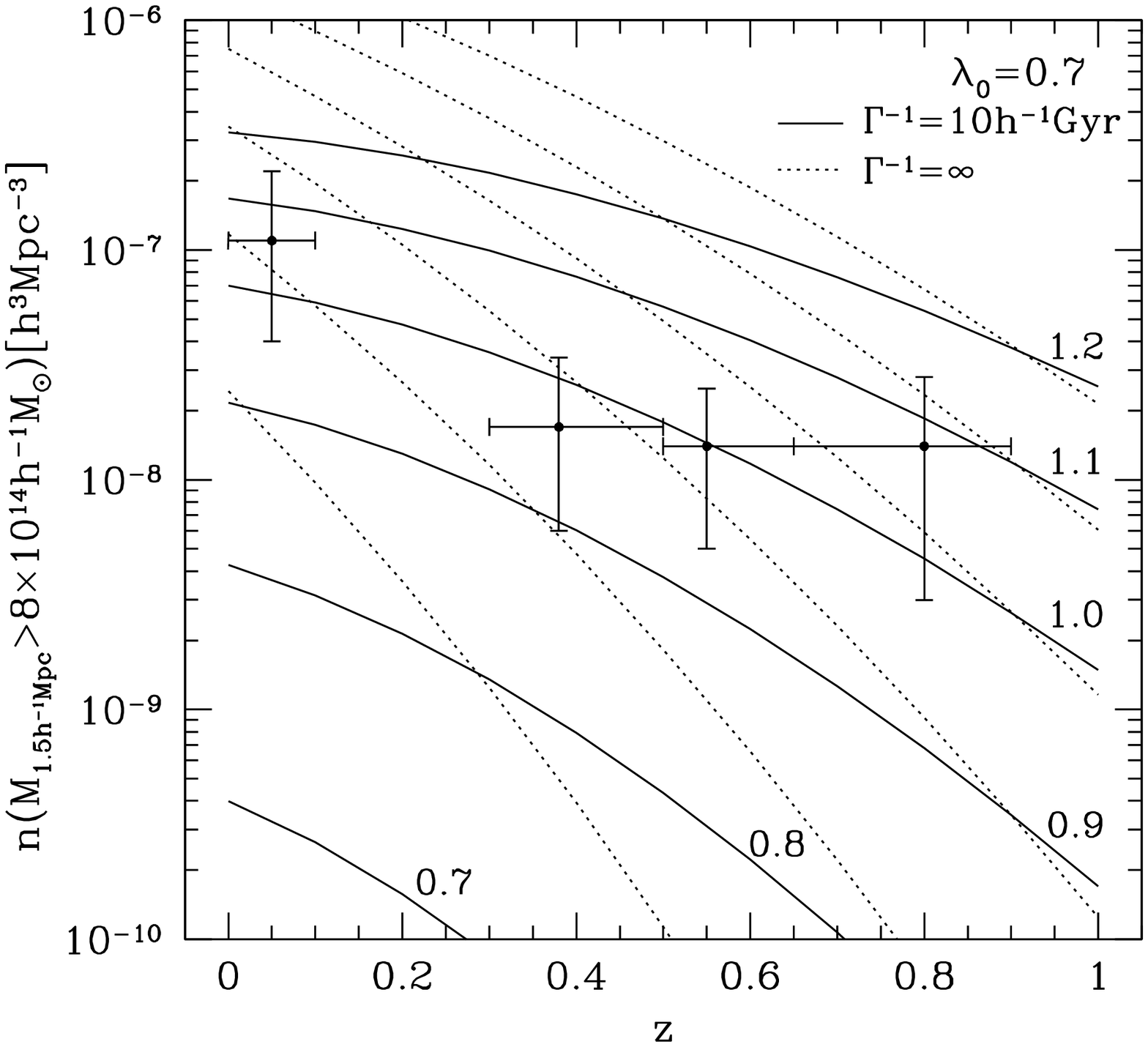}}}}
\figcaption{Evolution of the cluster abundance for $\lambda_0=0.7$. The
 number of clusters with mass 
 $M_{1.5h^{-1}{\rm Mpc}}>8\times 10^{14}h^{-1}M_\odot$, where 
 $M_{1.5h^{-1}{\rm Mpc}}$ is the mass within a comoving radius of
 $1.5h^{-1}{\rm Mpc}$, is plotted as a function of redshift. The filled
 circles with errorbars are observed abundance (see text for details). 
 Solid lines are predicted cluster abundance with lifetime of dark matter
 $\Gamma^{-1}=10h^{-1}{\rm Gyr}$, while we assume no decay for the
 dotted lines. We plot curves with $\sigma_8=0.7$, $0.8$, $0.9$, $1.0$,
 $1.1$, and $1.2$ (only solid lines are labeled). Since adopted mass
 function in this paper is based on the theoretical model of
 \citep{press74} which is now known to be inaccurate, curves may be
 different in more accurate calculations. In particular, our mass
 function may underpredict the large-mass halos, and this may result in
 faster decline of the curves with $z$.
\label{fig:cluabu}}
\vspace{0.5cm}

To see how large decay rate is required to explain the observations, in
Figure \ref{fig:gs} we show the constraints from low-$z$
\citep{bahcall03a} and high-$z$ \citep{bahcall03b} cluster abundances
assuming $\lambda_0=0.7$. We find that the lifetime of
$\Gamma^{-1}\lesssim 10h^{-1}{\rm Gyr}$ can explain both low-$z$ and
high-$z$ cluster abundances. Interestingly, this decay rate is roughly
consistent with that derived from the evolution of mass-to-light ratios
$\Gamma^{-1}\sim 10^{-1}{\rm Gyr}$ \citep{ichiki03}. 

\vspace{0.5cm}
\centerline{{\vbox{\epsfxsize=7.5cm\epsfbox{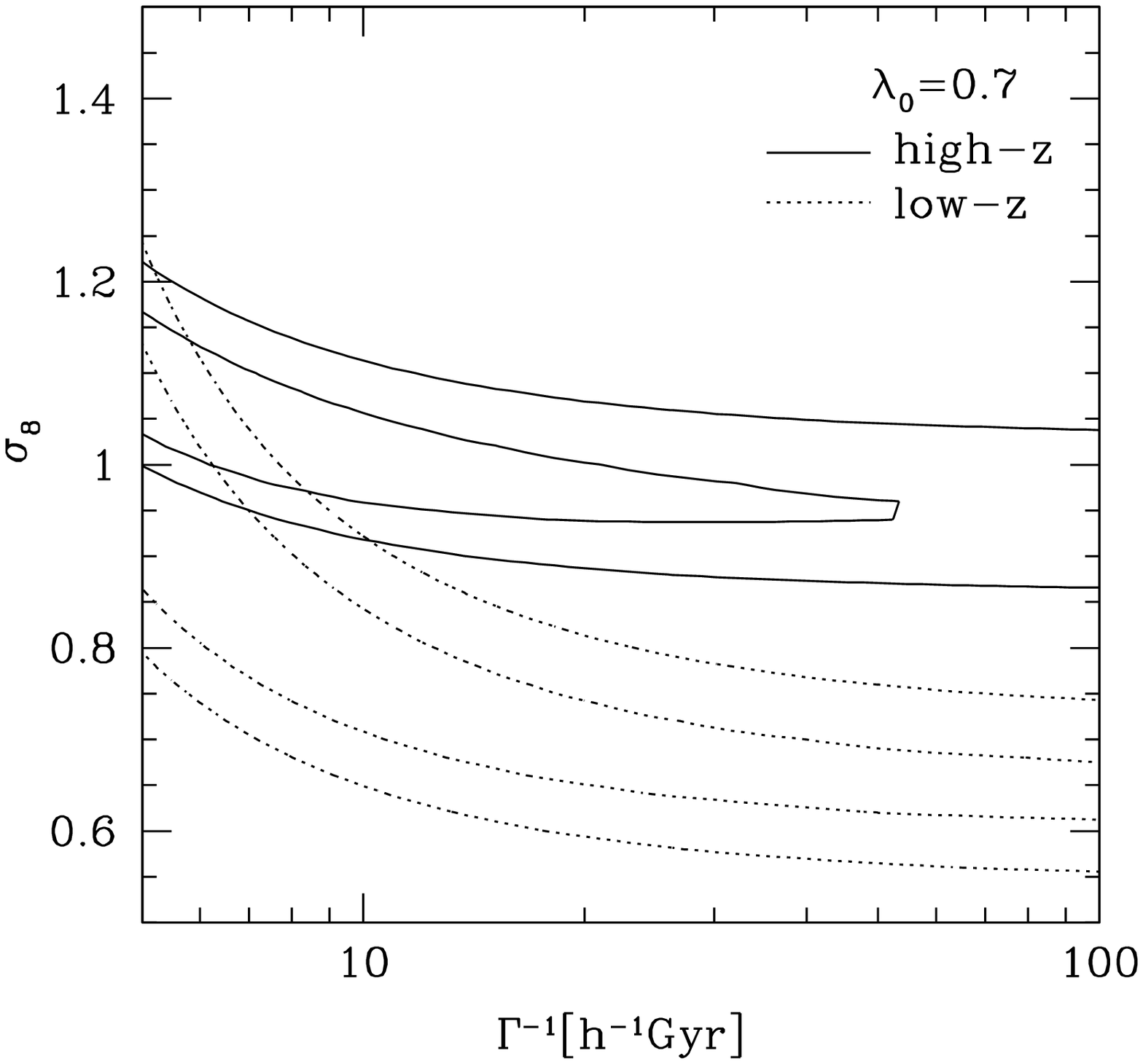}}}}
\figcaption{Constraints from high-$z$ ({\it solid}) and low-$z$ ({\it
 dotted}) cluster abundances in the $\Gamma^{-1}$-$\sigma_8$ plane.
 The observed abundance of high-$z$ clusters is taken from
 \citet{bahcall03b}, and that of low-$z$ clusters is from the SDSS
 clusters \citep{bahcall03a}.
 Contours of $68\%$ and $95\%$ limits are plotted.
 We adopt $\lambda_0=0.7$.
\label{fig:gs}}
\vspace{0.5cm}

We plot the constraint in the $(1-\lambda_0)$-$\sigma_8$ plane in
Figure \ref{fig:os} with $\Gamma^{-1}=7h^{-1}{\rm Gyr}$. We also show the
constraint from the observation of Type Ia supernovae complied by
\citet{tonry03} assuming the same value of $\Gamma^{-1}$. 
We use the sample of 172 supernovae with $z>0.01$ and
$A_V<0.5$ mag to avoid possible systematics. The resulting limit on
$1-\lambda_0$ is $1-\lambda_0=0.22^{+0.3}_{-0.4}$ (68\% confidence
limit), slightly lower than the limit in the case of no decay
($1-\lambda_0\sim 0.28$). We find that the best-fit parameter set
$(1-\lambda_0,\sigma_8)=(0.2,1.2)$ is consistent with the
result of Type Ia supernovae as well as the evolution of the cluster
abundance.  

\vspace{0.5cm}
\centerline{{\vbox{\epsfxsize=7.5cm\epsfbox{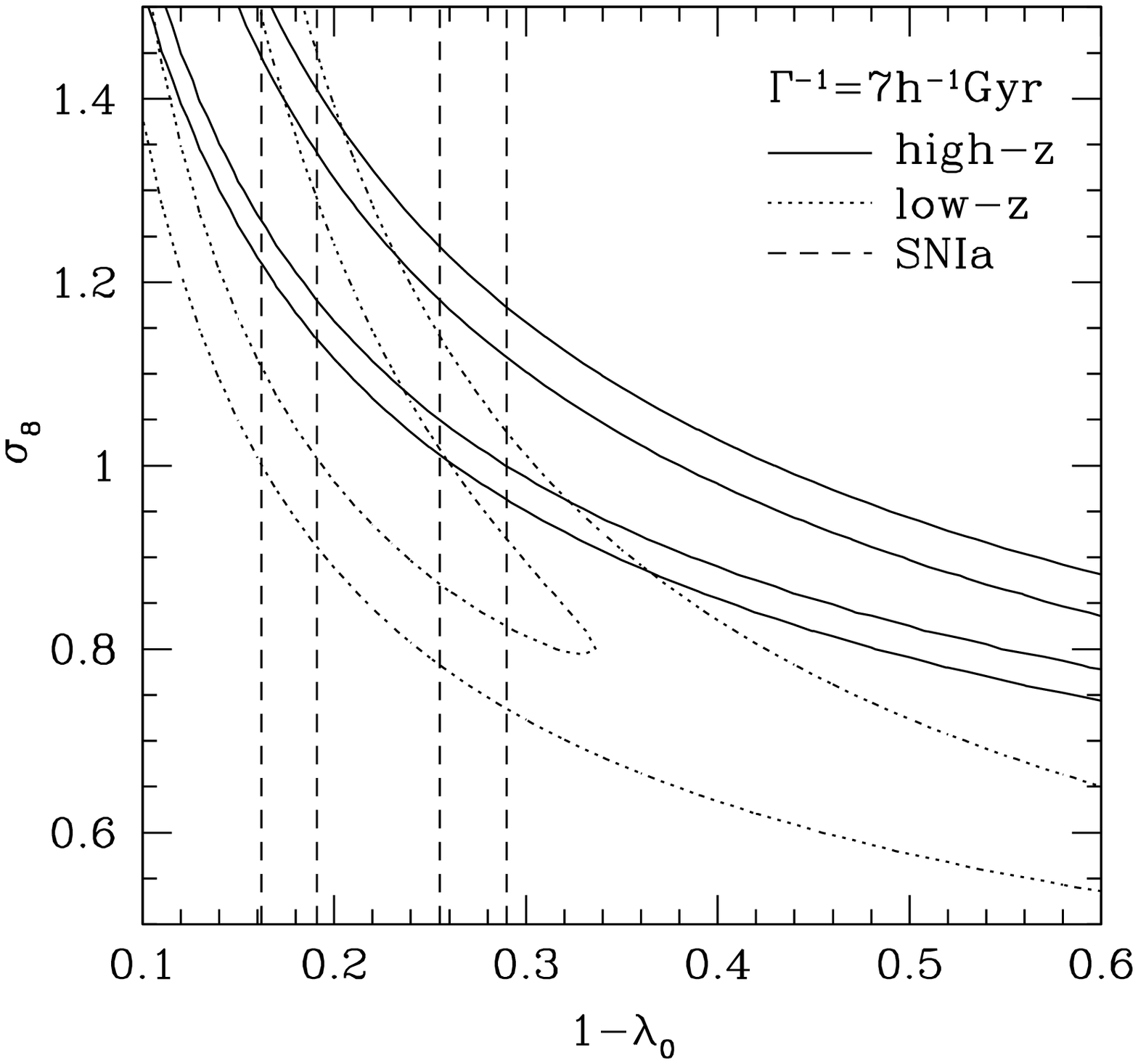}}}}
\figcaption{Constraints in the $(1-\lambda_0)$-$\sigma_8$ plane assuming
 $\Gamma^{-1}=7h^{-1}{\rm Gyr}$. We also plot the $68\%$ and $95\%$
 confidence limits from 172 Type Ia supernovae complied by \citet{tonry03}.
 The best-fit parameter set of the combination of these three
 constraints is $(1-\lambda_0,\sigma_8)=(0.2,1.2)$.
\label{fig:os}}
\vspace{0.5cm}

\section{Summary}

We have presented structure formation model in decaying cold dark matter
universe. More specifically, we have calculated the spherical model and
construct the mass function in the similar way as the theory of
\citet{press74}. Using this mass function, we have shown that decaying
cold dark matter model with $\Gamma^{-1}\lesssim 10h^{-1}{\rm Gyr}$ can
explain observed cluster abundances at both low-$z$ and
high-$z$ universe even better than the usual stable cold dark matter
model. Our model may be also consistent with the excess of cosmic
microwave background fluctuations on small scales; large mass
fluctuations $\sigma_8\sim 1.1$ are required to account for this excess
if it is interpreted as the Sunyaev-Zel'dovich effect of the cluster at
$z\sim 1$ \citep{komatsu02,bond03}.  

While we have concentrated on the evolution of the cluster abundance, it
is interesting to ask whether the decaying cold dark matter model is
consistent with other observations such as the anisotropy of cosmic
microwave background and large-scale structure of the universe. 
For instance, \citet{percival01} constrained the cosmological parameters
as $\Omega_{m0}h=0.20\pm0.03$ from the power spectrum of galaxy
clustering. Since the power spectrum mainly measures the epoch of the
matter-radiation equality, $\Omega_{m0}h$ should be interpreted as
$\Omega_{m0}e^{\Gamma t_0}h$ in the decaying cold dark matter model.
In the case of $h=0.7$, $\Gamma^{-1}=7h^{-1}{\rm Gyr}$, we constrained
$1-\lambda_0$ as $0.19<1-\lambda_0<0.23$ with 68\% confidence level.
This allowed range corresponds to $0.23<\Omega_{m0}e^{\Gamma
t_0}h<0.27$, thus marginally consistent with the observed power
spectrum. Moreover, the situation becomes better if the Hubble constant
is smaller than 0.7. For the anisotropy of cosmic microwave background,
it is difficult to infer the consequence of decaying cold dark matter
model because many parameters degenerate. We plan to study 
comprehensive constraints of these observations on decaying cold dark
matter model, which will be presented elsewhere.  

We note that our modeling of mass function may be inaccurate because it
has been claimed that the mass function of \citet{press74}, particularly
at high mass region, tends to underestimate the abundance of the 
massive halos compared with $N$-body simulations \citep[e.g.,][]{jenkins01}.
This may be a part of reason that the cluster abundance in our modeling
based on the theory of \citet{press74} seems to decline slightly faster
with redshift than that in \citet{bahcall03b} who used the result of
$N$-body simulations. Thus the constraints on the decay rate
$\Gamma$ (e.g., Figure \ref{fig:gs}) may be weaken a bit if we use
more accurate mass function in decaying cold dark matter model. One
way to refine the model is to resort $N$-body simulations, as in
the case of the standard cold dark matter model. Nevertheless we believe
that our qualitative result that decaying cold dark matter model tends
to lower the ratio of the local cluster abundance to the high-redshift
cluster abundance is still unchanged even if we adopt more accurate mass
functions. The cluster abundance at higher redshift universe $z\sim
1-2$, which will be obtained by future surveys of X-ray or
Sunyaev-Zel'dovich effects, can tightly constrain the decay rate of dark
matter.   

\acknowledgments
We thank Kiyotomo Ichiki for useful discussions.
K.T.'s work is supported by Grant-in-Aid for JSPS Fellows.


\end{document}